# Is a spectrograph of hidden variables possible?

**Alejandro A. Hnilo**

*CEILAP, Centro de Investigaciones en Láseres y Aplicaciones, (MINDEF-CONICET);*
*J.B. de La Salle 4397, (1603) Villa Martelli, Argentina.*
*email: alex.hnilo@gmail.com*

*Abstract.*

A new definition of "Realism" is proposed: it is that a gedanken "spectrograph" of hidden variables behaves as an actual (say, wavelength) spectrograph. The question is: does this definition allow, by itself, the derivation of Bell's inequalities? If it were, then such a spectrograph would be impossible, for Bell's inequalities are observed to be violated. In this short paper it is reported that, on the contrary, such spectrograph is compatible with the violation of Bell's inequalities. This result puts some new light on the controversy about the hypotheses necessary to derive Bell's inequalities. In particular, "Spectrograph's Realism", and "Locality", are proven to be different, and both necessary, hypotheses to derive Bell's inequalities.

*February 28th, 2023.*

## 1. Introduction.

Quantum Mechanics (QM) predicts the violation of Bell's inequalities, and experiments confirm this prediction. The inequalities are derived from intuitive ideas, which are usually shorthanded as "Locality" and "Realism". However, the precise definitions of these ideas are difficult and controversial. As a consequence, some scholars state that Locality and Realism mean essentially a single hypothesis, and that what is false is "Local Realism" [1,2]. Others argue that Quantum Mechanics (QM) is strictly Local [3-5], and that the violation of Bell's inequalities is a consequence of the wavy nature of matter [6]. It has also been claimed that only Realism is falsified by the violation of Bell's inequalities, and that Locality plays no role in the problem [7]. Independently, Bell's inequalities have been derived from the sole condition that the *series* of outcomes in station "A" ("B") is independent of the setting in "B"("A") [8]. This condition seems to be the ultimate form of *non-contextuality* [9]. On the other hand, the description of QM problems within a Boolean algebra (which is at the basis of the classical theory of probability) requires, in general, the definition of a Boolean sub-algebra, or *context* [10].

In this short paper, the consequences of a proposed new definition of Realism (the "hidden variables' spectrograph"), are explored. The result hopefully puts some new light on the meaning of the assumptions necessary to derive Bell's inequalities. In the next Section, the derivation of the Clauser-Horne (CH) inequality is reviewed, for it is useful for the discussion that comes next. In Section 3, the idea of the "hidden variables' spectrograph" is introduced, and its consequences are derived. The conclusion, in few words, is that Locality and Realism (as they are defined here) are indeed *separate* hypotheses, both necessary to derive Bell's inequalities.

## 2. Review of the derivation of the Clauser-Horne (CH) inequality.

It worth mentioning here that the CH inequality has been used in the loophole-free experiments with photons using Eberhardt's states [11-12]. Its detailed derivation can be found in [9]. Assume that the probability to detect a photon transmitted by a polarization analyzer oriented at an angle $\alpha$ in station A (Figure 1) is $P_A(\alpha,\lambda)$, where $\lambda$ is an arbitrary but classical, counterfactual definite, "hidden" variable. The observable probability of detection is a statistical average on $\lambda$:

$$P_A(\alpha) = \int d\lambda . \rho(\lambda) . P_A(\alpha,\lambda) \qquad (1)$$

where $\rho(\lambda) \geq 0$, $\int d\lambda . \rho(\lambda) = 1$ and $0 \leq P_A(\alpha,\lambda) \leq 1$. Besides, the integral is "well behaved" (= it is a Riemann or Lebesgue integral). The set of these assumptions is one of the possible definitions of "Realism". Note that classical probabilities are involved, so that a Boolean algebra at the hidden variables level is presupposed.

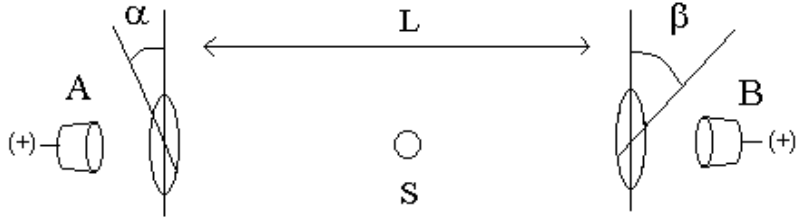

Figure 1: Sketch of an experiment to measure the CH inequality. Source S emits pairs of photons entangled in polarization towards stations A and B separated by a large distance L. At each station, a polarization analyzer is set at some angle $\{\alpha,\beta\}$. The probability of coincidences $P_{AB}(\alpha,\beta)$ is measured from the number of detections that occur simultaneously after both analyzers. The QM-predicted and measured $P_{AB}(\alpha,\beta)$ values violate the CH inequality.

Consider now two photons carrying the same value of $\lambda$. The probability that both are detected after analyzers set at angles $\{\alpha,\beta\}$ in stations A and B is, by definition, $P_{AB}(\alpha,\beta,\lambda)$. A usual form of "Locality" is the assumption that instantaneous action-at-a-distance effects do not exist, or that the experiments at A and B are somehow separable [13]. This assumption implies that:

$$P_{AB}(\alpha,\beta,\lambda) = P_A(\alpha,\lambda).P_B(\beta,\lambda). \tag{2}$$

what means *statistical independence* in the hidden variables' space. The observable probability of coincidences is then:

$$P_{AB}(\alpha,\beta) = \int d\lambda.\rho(\lambda).P_A(\alpha,\lambda).P_B(\beta,\lambda). \tag{3}$$

It is also assumed that $\{\alpha,\beta,\lambda\}$ are statistically independent variables: $P_A(\alpha,\lambda) = P_A(\alpha).P_A(\lambda)$, and that $\rho(\lambda)$ is independent of $\{\alpha,\beta\}$. The set of these assumptions is often named "non-contextuality" [9]. As eqs.2-3 involve classical probabilities, it is sometimes argued that (this form of) "Locality" is a hypothesis inseparable of "Realism".

Now, given $0 \leq x, x' \leq X$, $0 \leq y, y' \leq Y$, the following inequalities hold:

$$-XY \leq xy - xy' + x'y + x'y' - Xy - Yx' \leq 0 \tag{4}$$

choosing $x = P_A(\alpha,\lambda)$, $x' = P_A(\alpha',\lambda)$, $y = P_B(\beta,\lambda)$, $y' = P_B(\beta',\lambda)$ and $X=Y=1$, then:

$$-1 \leq P_A(\alpha,\lambda).P_B(\beta,\lambda) - P_A(\alpha,\lambda).P_B(\beta',\lambda) + P_A(\alpha',\lambda).P_B(\beta,\lambda) + P_A(\alpha',\lambda).P_B(\beta',\lambda)$$
$$- P_B(\beta,\lambda) - P_A(\alpha',\lambda) \leq 0 \tag{5}$$

where $\{\alpha,\beta,\alpha',\beta'\}$ are different angle settings at the stations (see Fig.1). Applying $\int d\lambda.\rho(\lambda)$ and eqs.1 and 3 to get observable probabilities:

$$-1 \leq P_{AB}(\alpha,\beta) - P_{AB}(\alpha,\beta') + P_{AB}(\alpha',\beta) + P_{AB}(\alpha',\beta') - P_B(\beta) - P_A(\alpha') \equiv J \leq 0 \tag{6}$$

which is the CH inequality. The QM predictions violate it. F.ex., for the Eberhardt's state $|\psi_E\rangle = (1+r^2)^{-\frac{1}{2}}.\{|x_A,y_B\rangle + r.|y_A,x_B\rangle\}$ ($r^2 \ll 1$, usually $r^2 \approx 0.1$ [11,12]), $P_{AB}(\alpha,\beta) = (1+r^2)^{-1}.[cos(\alpha).sin(\beta) + r.sin(\alpha).cos(\beta)]^2$, $P_A(\alpha') = (1+r^2)^{-1}[cos^2(\alpha') + r^2.sin^2(\alpha')]$ and $P_B(\beta) = (1+r^2)^{-1}[r^2.cos^2(\beta) + sin^2(\beta)]$.

Choosing the angle settings so that $cos(\alpha') \approx 0$ and $sin(\beta) \approx 0$, the single probabilities $P_B(\beta)$ and $P_A(\alpha')$ are $\approx r^2$; the coincidence probabilities $P_{AB}(\alpha,\beta)$, $P_{AB}(\alpha',\beta')$ and $P_{AB}(\alpha',\beta)$ are also $\approx r^2$. The settings $\{\alpha,\beta'\}$ are still free to make $P_{AB}(\alpha,\beta') \approx 0$. Then $J \approx 3r^2 - 2r^2 = r^2 > 0$, violating eq.6. Therefore, QM and experiments are incompatible with at least one of the hypothesis leading to eq.6. Nevertheless, as it was briefly discussed in the Introduction, the precise interpretation of these hypotheses has led to controversy.

## 3. Spectrograph's Realism.

Let rephrase the definition of Realism. Let assume that it is possible to think of a "hidden variables' spectrograph" able to record, in each of its channels (each one identified with the index $i$, or $i$-channel), the number of particles carrying the value $\lambda_i \pm \Delta\lambda/2$ of the hidden variable ($\Delta\lambda$ is the resolution of the spectrograph). Even though the histograms or "spectra" produced by such hypothetical device cannot be observed (the hidden variables are assumed unobservable), some of their would-be features can be known if the gedanken spectrograph behaves like an actual (say, wavelength) spectrograph. These features are:

1. *The number $N_i$ of particles detected in any i-channel is positive, or zero.*
2. *The sum of the $N_i$ over all the i-channels is the total number of detected particles, $\Sigma N_i = N$.*

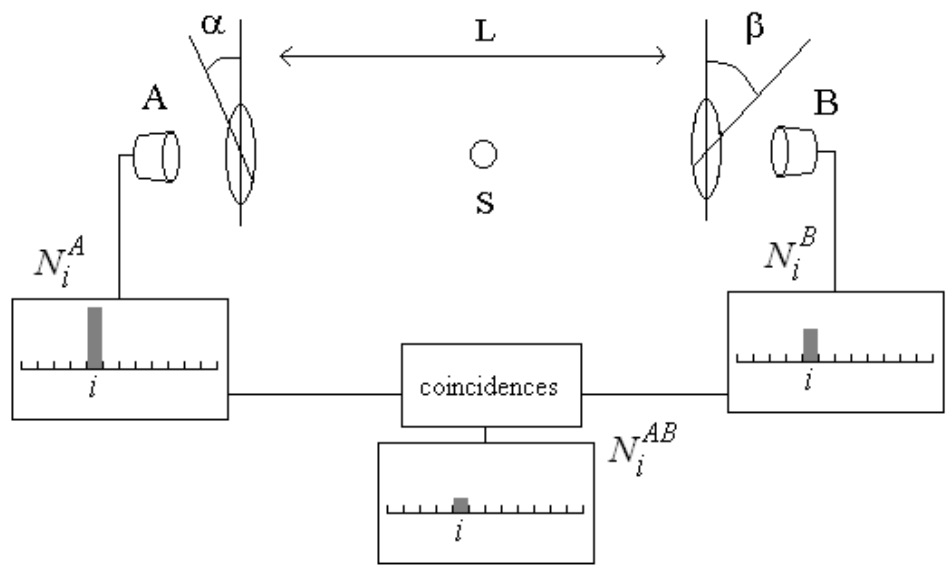

Figure 2: A "hidden variables' spectrograph" is placed in each station, which counts the number of detections for each value of the hidden variable (gray columns, here drawn for $i$=5 only), and also the time values when the detections occur ("time stamping"). The number of coincidences for each value of the hidden variable is then calculated and plotted below.

Let assume that identical hidden variables' spectrographs are placed in each station A and B (see Figure 2). Thus, $N^A_i$ ($N^B_i$) is the number of particles detected in each $i$-channel in the spectrograph in station A (B) (for clarity, the explicit dependence on the settings {α,β} is dropped here). Recall that the CH inequality involves detections after the "transmitted" gate of the analyzers only, the "reflected" ones are not taken into account. Otherwise, two spectrographs per station are necessary. The number of coincidences in channel $i$ is $N^{AB}_i$. The setup records not only the number of particles detected in each channel in each station, but also the time value when each particle is detected ("time stamping" [14,15]). The $N^{AB}_i$ are then calculated by counting how many particles were detected simultaneously in A and B for the same $i$-channel. Simultaneous detections occurring in different $i$-channels are considered noise and discarded, for valid entangled particles are assumed to carry the same value of λ. In these conditions:

3. $\forall i$, $N^{AB}_i \leq Minimum\{N^A_i, N^B_i\}$.

For, because of the way the $N^{AB}_i$ are calculated, there cannot be more coincidences than single counts in any $i$-channel. The set of features #1-#3 define "Spectrograph's Realism". In this definition of Realism, probabilities are not directly involved, only the number of detections that would-be recorded in the gedanken spectrographs.

In practice, the probability $P_A(α,λ)$ in eq.1 would be measured as the limit (for large $N$) of $N^A_i/N$, and the same for $P_B(β,λ)$ and $P_{AB}(α,β,λ)$, so that the eq.2 becomes:

$$N^{AB}_i / N = N^A_i \times N^B_i / N^2 \tag{7}$$

which implies the validity of #3. Note that eq.7 implies #3. Yet, the inverse implication is not true.

Be aware that a *non*-Boolean hidden variables' model [16] does not hold to #3. Besides, because of #1, Spectrograph's Realism is incompatible with probabilities defined outside the [0,1] interval [17], and hence, with contextuality [10]. In spite of these incompatibilities, Spectrograph's Realism is compatible with the violation of Bell's inequalities, as shown below.

The question now is: is it possible deriving the CH inequality (eq.6) from #1-#3 *only*? The answer is found after a simple calculation. From #3, for all $i$-channels:

$$N^{AB}_i (α,β) \leq N^B_i (β) \tag{8a}$$

$$N^{AB}_i (α',β) \leq N^A_i (α') \tag{8b}$$

Let define as $Γ_1$ the set of $i$-channels (= the values of λ) where:

$$N^{AB}_i (α',β') \leq N^{AB}_i (α,β') \tag{9a}$$

and $Γ_2$ its complement, that is, the set of $i$-channels where:

$$N^{AB}_i (α',β') > N^{AB}_i (α,β') \tag{9b}$$

Using eqs.8 and 9a, in each $i$-channel in $Γ_1$ (indicated with super index 1) is valid that:

$$N^{(1)AB}_i(α,β) + N^{(1)AB}_i(α',β) + N^{(1)AB}_i(α',β') - N^{(1)AB}_i(α,β') - N^{(1)A}_i(α') - N^{(1)B}_i(β) \leq 0 \tag{10}$$

For the first term is smaller or equal than the last one, the second one is smaller or equal than the fifth one, and the third one is assumed (in $Γ_1$) to be smaller or equal than the fourth one. In the set

$\Gamma_2$ instead, the corresponding equation is:

$$N^{(2)AB}_i(\alpha,\beta) + N^{(2)AB}_i(\alpha',\beta) + N^{(2)AB}_i(\alpha,\beta') - N^{(2)AB}_i(\alpha',\beta') - N^{(2)A}_i(\alpha') - N^{(2)B}_i(\beta) < 0 \quad (11)$$

For the fourth term is assumed (in $\Gamma_2$) to be larger than the third one (see eq.9b, the other terms are related as in eqs.8). Now sum up eqs.10 and 11, also sum and rest both $N^{(2)AB}_i(\alpha',\beta')$ and $N^{(2)AB}_i(\alpha,\beta')$, and rearrange terms. The notation of terms with the same setting in both sets $\Gamma_1$ and $\Gamma_2$ is contracted, f.ex.: $N^{(1)AB}_i(\alpha,\beta) + N^{(2)AB}_i(\alpha,\beta) = N^{AB}_i(\alpha,\beta)$, so that:

$$N^{AB}_i(\alpha,\beta) + N^{AB}_i(\alpha',\beta) + N^{AB}_i(\alpha',\beta') - N^{AB}_i(\alpha,\beta') - N^{A}_i(\alpha') - N^{B}_i(\beta) +$$
$$+ 2\times N^{(2)AB}_i(\alpha,\beta') - 2\times N^{(2)AB}_i(\alpha',\beta') < 0 \quad (12)$$

Now sum up over all *i*-channels to get actually observable numbers of detections. Note that the resulting inequality involves numbers of detections only, so that it would directly apply to experimental data. In order to allow comparison with the CH inequality, divide all terms by the total number of detected particles $N \gg 1$ (as usual, $N$ is assumed the same for all angle settings) to get:

$$P_{AB}(\alpha,\beta) - P_{AB}(\alpha,\beta') + P_{AB}(\alpha',\beta) + P_{AB}(\alpha',\beta') - P_B(\beta) - P_A(\alpha') +$$
$$- (2/N)\times\Sigma_i\{N^{(2)AB}_i(\alpha',\beta') - N^{(2)AB}_i(\alpha,\beta')\} < 0 \quad (13)$$

The first line in eq.13 is J as defined in eq.6, so that the inequality derived from Spectrograph's Realism finally reads:

$$J < (2/N)\times\Sigma_i\{N^{(2)AB}_i(\alpha',\beta') - N^{(2)AB}_i(\alpha,\beta')\} \quad (14)$$

The precise value of the rhs is unknown, for the features of the hidden variables' space are unknown by definition. But it is definite positive, because $N^{AB}_i(\alpha',\beta') > N^{AB}_i(\alpha',\beta') \; \forall i \in \Gamma_2$ (see eq.9b). Therefore, unless the measure of the set $\Gamma_2$ is zero, the usual CH inequality ($J \leq 0$) does not hold. Eq.14 is therefore compatible with both QM predictions and experimental results. Of course, the CH inequality is retrieved if the validity of eq.7 is assumed.

**Conclusion.**

It is demonstrated that the hypothesis of "Spectrograph's Realism" (features #1-#3) is compatible with the violation of the CH inequality. This result strengthens the importance of statistical independence (here, eq.7) in the derivation of Bell's inequalities. For, in logical terms, a counterfactual definite, non-contextual and statistically *not*-independent hidden variables model able to reproduce the observed violation of Bell's inequalities is possible. Spectrograph's Realism and eq.7 are hence separate, and both necessary, hypotheses to derive the CH inequality. Note that these hypotheses circumvent the use of classical probabilities; they only involve numbers of detections.

**Acknowledgements.**

This work received support from the grants N62909-18-1-2021 Office of Naval Research Global (USA), and PIP2022-0484CO from CONICET (Argentina).